# Interfacial transport driven by electrohydrodynamic instability at the plasma-liquid interface

**Seungjun Lee [1,2], Woojin Nam [3,4] and Gunsu Yun [2,4*]**
[1] Pohang Accelerator Laboratory, Pohang University of Science and Technology, Pohang 37673, Republic of Korea.
[2] Division of Advanced Nuclear Engineering, Pohang University of Science and Technology, Pohang 37673, Republic of Korea
[3] Mechatronics Research, Samsung Electronics Co., Ltd, Hwaseong 18448, Republic of Korea
[4] Department of Physics, Pohang University of Science and Technology, Pohang 37673, Republic of Korea
*E-mail: gunsu@postech.ac.kr

**Abstract**

Interfacial dynamics play a central role in transport processes across the plasma–liquid interface. While the strong electric field in the plasma sheath can destabilize the liquid surface and induce species transfer from the liquid into the plasma, the mechanistic relationship between surface instability and interfacial transport remains poorly understood. Here, we investigate the transport of sodium species in an atmospheric-pressure helium plasma in contact with a negatively DC-biased NaCl electrolyte using high-speed imaging, laser Mie scattering, and optical emission spectroscopy. The results show that Na transport is mediated by droplet emission from the liquid surface and proceeds through three sequential stages: surface deformation, Taylor cone formation, and electrospray. The threshold voltage required for Na I optical emission decreases with decreasing surface tension, in good agreement with the marginal condition for electrohydrodynamic (EHD) instability predicted by linear perturbation analysis. These findings demonstrate that EHD-driven droplet emission is the primary mechanism carrying sodium ions from the liquid into the plasma, where they are neutralized and excited. More broadly, this work establishes the surface instability as the active transport channel that governs the injection of dissolved species in the liquid into the plasma, creating a unique reaction network of plasma chemistry.

**Introduction**

Transport across the plasma-liquid interface plays a crucial role in plasma-liquid interactions, which represent a unique convergence of physical and chemical processes, distinctly different from those found in solid–liquid or gas–liquid systems [1, 2]. At the interface, steep gradients in temperature, density, and electric field create a highly dynamic, nonequilibrium environment in which species and energy transport play an important role. Interfacial transport—encompassing phenomena such as particle adsorption, desorption, sputtering, and electrolysis—strongly influences plasma-induced chemistry in both the plasma phase and the liquid bulk [1, 3]. As a result, interfacial transport has garnered significant attention in diverse fields, including analytical chemistry, nanoparticle synthesis, microfluidics, electrolysis, advanced oxidation engineering, and plasma medicine [4-8]. However, the intrinsically complex feedback loops and dynamic nature of the sheath-liquid interface present significant challenges in elucidating the underlying mechanisms of transport across this boundary [1, 9, 10].

One of the most critical factors governing interfacial transport is the instability of the liquid surface induced by electromagnetic fields and plasma flow. Such instabilities at plasma-facing liquid surfaces arise from the competition between electrostatic and fluid-dynamic forces, and can lead to surface deformation, droplet ejection, and the entrainment of liquid-phase species into the plasma. In atmospheric-pressure plasma-liquid systems, this instability-driven transport has direct consequences for plasma chemistry, since microdroplets and solute species carried from the liquid into the plasma phase can profoundly alter the local chemical composition and reaction network of the plasma.

The entrainment of microdroplets and solute atoms from the liquid phase into the plasma is not merely a passive transport event—it can substantially alter the plasma chemistry network itself. Microdroplets introduced into an atmospheric-pressure plasma act as localized sources of water vapor and dissolved species, generating steep gradients in reactive oxygen and nitrogen species (RONS) within the plasma volume. Wang et al. [11] demonstrated using laser-induced fluorescence that even a single microdroplet (~41 μm) in a helium plasma produces significant local gradients in water vapor and OH radical concentrations, and that multiple droplets can cause local quenching of the discharge through water vapor accumulation. Beyond modifying radical densities, microdroplets entrained in the plasma function as microscale chemical reactors, where the coupling of charged droplets with afterglow species and (V)UV radiation substantially enhances interfacial reactivity relative to conventional bulk liquid interfaces [12], with reaction enhancement factors exceeding $10^3$ reported for plasma–microdroplet fusion platforms [13]. The charged state of electrospray-generated droplets further modifies their internal chemistry: in highly charged water droplets, the thermodynamics of redox reactions can be dramatically altered relative to the bulk, enabling spontaneous generation of hydroxyl and hydrogen radicals and solvated electrons without external activation [14]. The transport of dissolved solutes into the plasma via droplet emission adds a further dimension to this chemistry. In solution-cathode glow discharge systems, it is well established that dissolved elemental species transferred from the liquid cathode into the plasma are excited and detected by atomic emission spectroscopy [15], confirming that species transported across the plasma-liquid interface directly participate in gas-phase reaction pathways. Collectively, these findings indicate that EHD-driven droplet emission represents a chemically active interfacial process: the composition, charge state, and size distribution of ejected droplets collectively determine which solute species enter the plasma-phase reaction network and how they perturb it.

Recently, similar instability-induced transport phenomena have been reported in atmospheric pressure plasma-liquid systems. In applications such as electrolysis and nanomaterial synthesis, the transfer of metal species from the liquid phase to the plasma has been observed, often attributed to droplet formation at deformed liquid surfaces. For instance, Shirai et al. [3] reported the emission of sodium atoms from an electrolyte into the plasma and proposed that the transport mechanism involves the release of metal atoms from liquid microdroplets. These droplets are believed to originate from the tip of a conically shaped liquid surface, distorted by the sheath electric field. When the electrostatic force exceeds the restoring surface tension, droplets are emitted from the tip of the Taylor cone—a process commonly referred to as electrospray or electrojet [16-18]. While electrohydrodynamic (EHD) instability at the sheath-liquid interface with plasma flow has been studied in both continuous fusion plasma and atmospheric plasma systems, a detailed understanding of these complex regions remains elusive [19-21]. A clear understanding of instability at plasma-facing liquid surfaces is essential for both stable plasma operation and precise control of plasma-liquid interactions.

The EHD instability of conducting liquid surfaces under strong electric fields is not merely a laboratory phenomenon. Taylor's original development of cone theory was explicitly motivated by the behavior of water droplets under thunderstorm electric fields [21], and the Taylor cone – electrospray sequence identified in those studies is the same physical process underlying the present work. More broadly, whenever a plasma or strongly electrified gas comes into direct contact with a conducting liquid—as occurs at the ocean surface beneath thunderstorms, or at the plasma-liquid interface formed by a lightning strike on water—the EHD instability mechanism governs the ejection of charged, surface-enriched microdroplets from the liquid into the gas phase. Recent experimental studies have shown that such plasma-liquid interfaces formed by lightning-like discharges act as electrochemical reactors, locally generating reactive species and bioavailable organic compounds at yields substantially exceeding those of gas-phase discharges alone [22, 23]. The selective transport of dissolved ionic and surface-active species from the liquid into the plasma via EHD-driven droplet emission, as elucidated in the present study, may contribute to the efficiency of these plasma-liquid coupled reactions by exposing dissolved species to the reactive plasma environment in a highly dispersed, microdroplet form. Despite its potential significance in both engineered and natural plasma-liquid systems, the microscopic mechanism by which dissolved species are transferred across a conducting liquid surface under electric forcing has not been directly visualized or quantitatively linked to EHD stability theory—a gap this work addresses.

In this study, we investigate the Na atom transport induced by surface instability during a plasma-liquid interaction under atmospheric pressure. We observed the optical emission of Na atoms in microwave-driven plasma in contact with negatively DC-biased electrolytes, along with the distortion in the liquid surface and micro-droplet spraying. To elucidate the mechanism of Na transport from liquid to plasma, we conducted high-speed imaging at the liquid surface and examined the DC bias voltage required for optical Na I emission under different conditions of electrolytes. From the experimental results, the optical Na I emission was observed to be accompanied by droplet emission, and the process of droplet production was described in three steps: (1) surface deformation induced by electric wind, (2) Taylor cone formation, and (3) electrospray. In the following, we provide physical interpretations for each step with high-speed imaging results to understand the deformations on the surface. Furthermore, we discuss the relation between the required voltage for optical Na I emission and EHD stability of the interface based on linear perturbation analysis of Holgate *et al.* [15].

## Method

The experimental configuration, as described in Fig. 1, consists of a microwave (MW) driven plasma and a negatively DC-biased NaCl solution. The MW discharge is facilitated by a coaxial transmission line resonator (CTLR), operating at a resonant frequency of 2.45GHz. Pure helium gas with a purity of 99.99% is fed toward the liquid surface through the coaxial nozzle of the CTLR, maintaining a flow rate of 1 standard liter per minute (SLM) using a mass flow controller. NaCl solutions ranging from 1 to 1000 mM, poured into a transparent quartz dish, served as a liquid cathode. The distance between the CTLR and the liquid surface is kept to 5mm. The outer electrode of CTLR is electrically grounded, while a negative DC voltage is applied to a graphite rod immersed in the solution by using a DC power supply (AONETECH, AP10002) in the constant voltage mode. The potential difference between the CTLR and graphite rod is measured using a high-voltage probe (Tektronix, P6015A), and the current is calculated from the voltage drop at a 50-ohm serially connected resistor. Note that the MW discharge, pre-implemented before applying DC bias voltage, could provide a stable transition in DC-discharge mode. By supplying an abundant charge into the gap between the CTLR and liquid cathode, MW discharge reduces the breakdown voltage required for DC channeling and prevents a rapid change in voltage-current characteristics during the ignition. As a result, this MW-DC combined configuration enhances the clarity and precision in the analysis of the discharge behavior, allowing for more accurate observations and measurements.

A high-speed camera (Photron, SA4 500K-C1) records spatiotemporal changes in the liquid surface with varying frame rates, utilizing a microscope objective lens (OLYMPUS, LMPlanFL 10x). The camera is placed with 50 degree of incident angle to the liquid surface. To investigate the correlation between optical Na emission and changes in the liquid surface, we adopt laser Mie scattering technique with a custom blue laser at a wavelength of 442 nm, which allows the visualization of the droplet emission at the liquid surface. A combination of a photo-multiplier tube and an optical band pass filter is employed to trigger the high-speed camera by sensing Na I line emissions at 589.0nm ( $^2P_{3/2} - {}^2S_{1/2}$) and 589.6 nm ( $^2P_{1/2} - {}^2S_{1/2}$) [24].

**Results**

3.1. V-I characteristics and optical emission of Na

Snapshots in Figure 2 show the negatively DC-biased MW plasma using a 100mM NaCl solution as the liquid cathode and illustrate the morphological changes at different bias voltage levels. As can be seen in the snapshots, a diffusive plasma channel was observed between the microwave plasma and the liquid cathode. As the bias voltage increased, filamentary bright channels were observed along with the distortion of the liquid surface. Above an applied voltage of -550V, optical Na emission was observed, providing direct evidence for the transport of Na that was initially dissolved in the liquid cathode. To explore the correlation between bias voltage and the optical emission of Na, we conducted experiments to measure the threshold voltage for the optical Na emission.

Figure 3 presents the bias voltage required for optical Na emission and the voltage-current characteristics obtained at various concentrations of NaCl liquid cathode. In general, ions in an electrolyte play a role in carrying the electric current, so the electrical conductivity of solution increases with higher ion concentration. However, when ions are present in excess, they can become saturated and no longer contribute to increased electrical conductivity due to interaction between ions or the nature of the solution. As a result, the voltage-current curves plateaus to a constant voltage level, as shown in Figure 3. Similar to the voltage-current characteristics, the threshold voltage for optical Na emission decreases with increasing NaCl concentration and plateaus to a constant voltage level. The plateau in the threshold voltage can be attributed to the convergence of the solution resistivity as saturation of ion concentration, which means that the voltage required for optical Na emission does not depend on the NaCl concentration. Assuming that the solution is conducting liquid with negligible resistance at high NaCl concentrations, and that the bias voltage can be approximated as a potential drop across the plasma, we can speculate that the transport of Na atoms is driven by an electrodynamic feature of the plasma-liquid interface.

3.2. Visualization of surface deformation

High-speed imaging was performed to directly observe the deformation of the liquid surface. As illustrated in Figure 4, a depression of the liquid surface was observed after the ignition of DC discharge, even though the optical emission of plasma became dimmer and was no longer visible due to the short exposure time of imaging. Note that impinging He gas flow with the microwave-driven plasma did not induce depression of the liquid surface in the given gap distance, suggesting that DC biased discharge channel exerted force on the liquid surface. We hypothesized that an EHD force induced by the drift motions of the charged particles in DC discharge channel produced pressure on the liquid surface, commonly called electric wind. Given that the liquid surface was negatively charged, the positive ions were accelerated toward the liquid cathode during the formation of the DC discharge channel. In the discharge channeling process, the momentum transfer from ions to neutrals could cause the electric wind.

Following the surface depression, a rim-shaped protrusion was observed (see Figure 5). When the bias voltage exceeded the threshold voltage for optical Na emission, Taylor cones formed on the protruded surface, accompanied by optical Na emission (see Figure 5). To visualize the droplet emission from the tip of Taylor cone, laser Mie-scattering imaging was performed. After the rim-shaped protrusion, liquid droplets were sprayed from the tip of the Taylor cone, as illustrated in the figure by a blue light emitter smaller than $10\ \mu$m (see Figure 6 and 7). Experimental observations provide strong evidence that the droplet generation via the electrospray facilitates the transfer of Na particles to the plasma region.

3.3. Correlation between threshold voltage for optical Na emission and surface tension

To achieve a better understanding of the correlation between droplet emission and Na transport, we measured the threshold voltage for optical Na emission at different liquid cathodes with varying surface tension. As shown in Figure 8, the reduction in surface tension leads to the lowering of the threshold voltage. Notably, despite the reduction in the bias voltage at which sodium optical emission occurs, the optical emission process was consistently accompanied by droplet spraying, directly observable through the laser Mie-scattered signal. In principle, the formation of Taylor cone induced by the electric filed and the subsequent breakup of its tip into small droplets are EHD processes governed by the multiple forces [18, 25-27]. Thus, the reduction of surface tension, acting as an internal force providing resilience of surface, facilitates the surface deformation by the electric field. In other words, the correlation between surface tension and threshold voltage of optical Na emission demonstrates that droplets emission induced by EHD surface instability is the direct cause of Na transport.

EHD stability analysis of the plasma-liquid interface enhance the understanding of marginal condition for the stability of the liquid surface. A theoretical account of the dynamics of the plasma-liquid interface has been provided by Holgate et al. [19]. The study described the linear analysis of the dispersion and stability of EHD surface waves, including the behavior of a plasma sheath near the liquid surface. The pressure jump condition on the liquid surface is given by

$$p_s\,\hat{n} + \hat{n}\cdot\overleftrightarrow{\sigma} - \gamma(\nabla\cdot\hat{n})\,\hat{n} = 0 \tag{1}$$

where $\hat{n}$ is the surface unit normal which points into the plasma region, $\overleftrightarrow{\sigma}$ is the electrical stress tensor at the sheath, and $-\gamma(\nabla\cdot\hat{n})$ is the Young-Laplace pressure determined by the surface tension $\gamma$ and the local surface curvature, $\kappa = -\nabla\cdot\hat{n}$. Note that the liquid system being studied was assumed to be conducting, inviscid, and incompressible. The stress tensor $\overleftrightarrow{\sigma}$ at the plasma sheath can be expressed by [19]

$$\overleftrightarrow{\sigma} = \varepsilon_0\mathbf{E}\mathbf{E} - \frac{\varepsilon_0 E^2}{2}\overleftrightarrow{\mathrm{I}} - m_i n_i \mathbf{u}_i\mathbf{u}_i - n_{e,s}k_B T_e \exp\left(\frac{e\phi}{k_B T_e}\right)\overleftrightarrow{\mathrm{I}} \tag{2}$$

where $\varepsilon_0$ is the vacuum permittivity, $\mathbf{E}$ is the electric field vector, and $\overleftrightarrow{\mathrm{I}}$ is the identity tensor. Furthermore, $m_i$ is the ion mass, $n_i$ is the ion density, $\mathbf{u}_i$ is the ion drift velocity, $n_{e,s}$ is the electron density at the sheath edge, $k_B$ is the Boltzmann constant, $T_e$ is the electron temperature, $e$ is the elementary charge, and $\phi$ is the electric potential. The first two terms on the right-hand side correspond to the electrostatic Maxwell stress, the third term is the ion ram pressure, and the last term is the electron thermal pressure. By linearizing Eq. (1) with a wave-like perturbation characterized by a wavenumber k, the marginal conditions for EHD instability can be obtained. Here, $k$ denotes the perturbation wavenumber and $\lambda_D$ is the Debye length, which characterizes the electrostatic shielding distance in the plasma. For collision-less cold plasma, the perturbation analysis shows that short wavelength instabilities ($k\,\lambda_D \gtrsim 1$) behave according to the conventional EHD theory, $\varepsilon_0 E_0^2 > \gamma k$, while the growth of long wavelength perturbations ($k\,\lambda_D < 1$) are suppressed by the ion ram pressure on the liquid surface [26]. Here, $E_0$ denotes the equilibrium electric field strength at the interface. For a highly collisional sheath in contact with a high-voltage liquid cathode, the ion ram pressure and electron thermal pressure can be ignored. Assuming the ion motion is driven by the mobility of highly collisional sheath, $\mathbf{u}_i \approx \mu_i\mathbf{E}$, and the wall potential is very large compared to electron temperature, $|e\phi| \gg k_B T_e$, the pressure on the surface can be simplified as

$$p_s = \left(\frac{m_i n_i \mu_i^2}{\varepsilon_0} - 1\right)\frac{\varepsilon_0}{2}E_s^2 + \gamma\kappa \tag{3}$$

where $\mu_i$ is the electrical mobility of ion. For atmospheric-pressure plasmas like our system, the collision frequency of ion-neutral momentum transfer $(\nu_i)$ is much larger than the ion plasma frequency $(f_{\mathrm{pi}})$, yielding $(m_i\,n_i\,\mu_i^2)/\varepsilon_0 \ll 1$. As a result, the marginal stability condition for the liquid surface is given as

$$E_0 = \sqrt{\gamma k/\varepsilon_0} \tag{4}$$

which is the same with the conventional criterion of EHD stability for liquid-vacuum interface. The stability limit in the equation provides additional insight into the surface tension dependence of the threshold voltage for optical Na emission. As can be seen in Figure 8, the measured threshold voltages lie on a line with the fitted curve $\mathrm{V_{th}} \sim \gamma^{0.47}$, closely approaching the stability limit $E_0 \sim \gamma^{1/2}$ predicted from the linear perturbation theory.

According to numerical analysis by Holgate et al. [19], the stable-to-unstable transition limit for an initially flat liquid surface requires a few thousands of volts in a typical atmospheric pressure plasma with electron temperature of 1 eV and density of $10^{19}$ m$^3$. On the other hand, in our study, the surface instability was observed at a few hundreds of volts, below the predicted stability limit. We speculate that the morphological deformation induced by the electric wind on the liquid surface reduces the electric potential for the transition. As shown in Figure 7, Taylor cones form on the crest of the liquid ridge, suggesting that geometric effects can reduce the voltage required for droplet emission in our system, and thus optical emission could be observed at a relatively low voltage, around 500V.

Based on these experimental observations and analyses, we conclude that sodium transport in atmospheric-pressure plasma–liquid systems is primarily governed by EHD instability at the liquid interface. The three-stage transport process—surface deformation, Taylor cone formation, and electrospray—was consistently observed under varying electrolyte conditions, indicating a robust and reproducible mechanism. Moreover, the dependence of the threshold voltage for Na I emission on surface tension is in good agreement the marginal stability condition predicted by linear perturbation theory. Together, these findings establish a direct mechanistic link between interfacial instability and atomic emission, demonstrating that electric field-induced droplet ejection is the primary pathway by which dissolved metal species are transferred from the liquid into the plasma. More broadly, this work identifies liquid-surface instability as a chemically active transport mechanism that governs the injection of solvated species into plasma-phase reaction networks, providing a foundation for the rational design and control of plasma–liquid interfaces in both engineered and natural environments.

## Conclusion

In this study, we demonstrated that the droplet emission induced by electrohydrodynamic (EHD) instability of liquid surface is the primary mechanism responsible for Na transport from the liquid phase into the plasma. Using high-speed camera and Laser Mie scattering system, we elucidated the mechanism of droplet generation at the surface of liquid cathode. Experimental observations revealed that the droplet production proceeds through three sequential stages: surface deformation, Taylor cone formation, and electro-spray. Notably, reducing the surface tension of liquid cathode lowers the threshold voltage required for optical Na emission, in good agreement with the marginal condition for EHD instability predicted by linear perturbation analysis of the interfacial pressure balance. When the electric field at the liquid surface exceeds this stability threshold a Taylor cone forms and charged droplets are emitted from its tip. These droplets transport dissolved sodium species into the plasma, where they are atomized and excited, resulting into the observed optical Na I emission.

More broadly, this work establishes that liquid-surface instability in plasma–liquid systems is not merely a hydrodynamic phenomenon, but a chemically active transport mechanism that governs the transfer of dissolved species across the plasma–liquid interface. Because the liquid electrode is deformable, plasma-generated electric fields and momentum flux can either destabilize or stabilize the interface depending on operating conditions [19]. Understanding this interplay provides fundamental insight into interfacial transport and plasma–liquid coupling. These findings are expected to advance plasma engineering by enabling improved control over species transfer, plasma chemistry, and discharge stability in applications including analytical chemistry, electrochemical synthesis, and plasma-assisted materials processing.

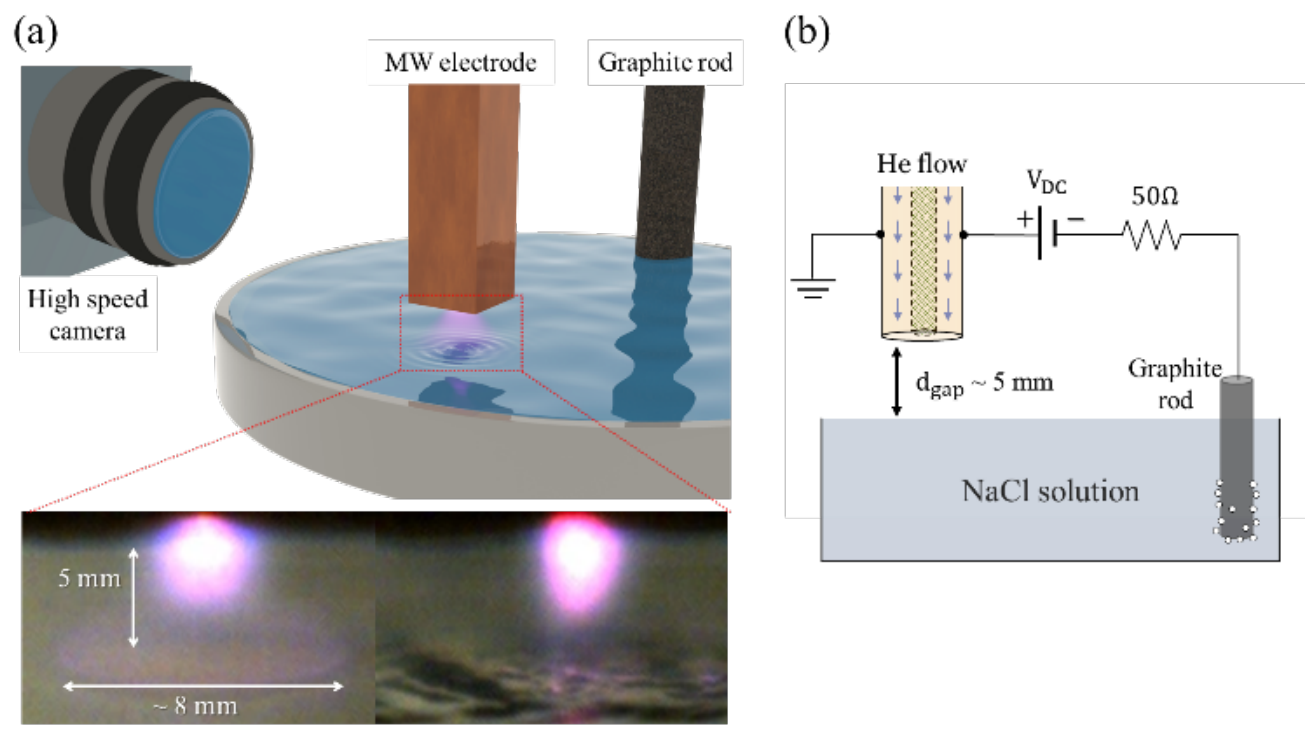

**Figure 1.** Schematic diagram of experimental setup consisting of (a) high-speed imaging system and (b) plasma-liquid interaction system. An electrically grounded coaxial transmission line resonator is located 5 mm above the liquid surface. The counter electrode, a graphite rod, is about 15 cm away from the plasma discharge and immersed in the solution. The high-speed camera is tilted at angle of 40 degrees to the horizontal liquid surface.

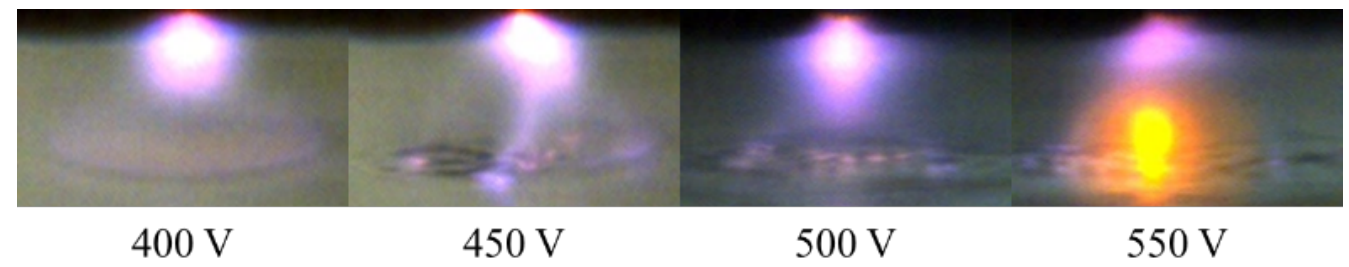

**Figure 2.** Snapshots of the plasma-liquid interaction by using High-speed camera. Note that concentration of NaCl solution is 100mM.

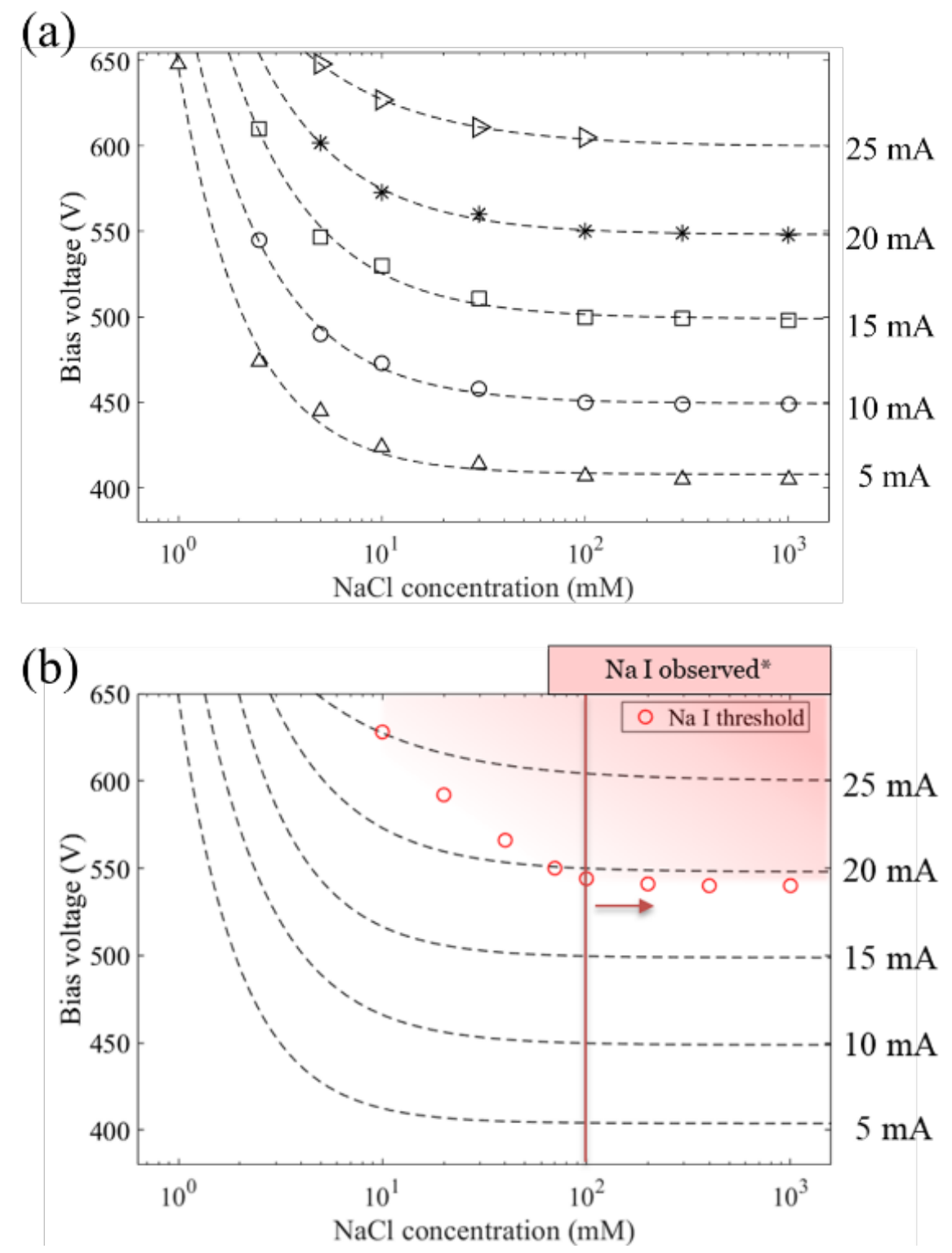

**Figure 3.** Bias voltage as a function of NaCl concentration at different current I = 5–45 mA. The red-colored region indicates the condtion in which Na I line mission is observed.

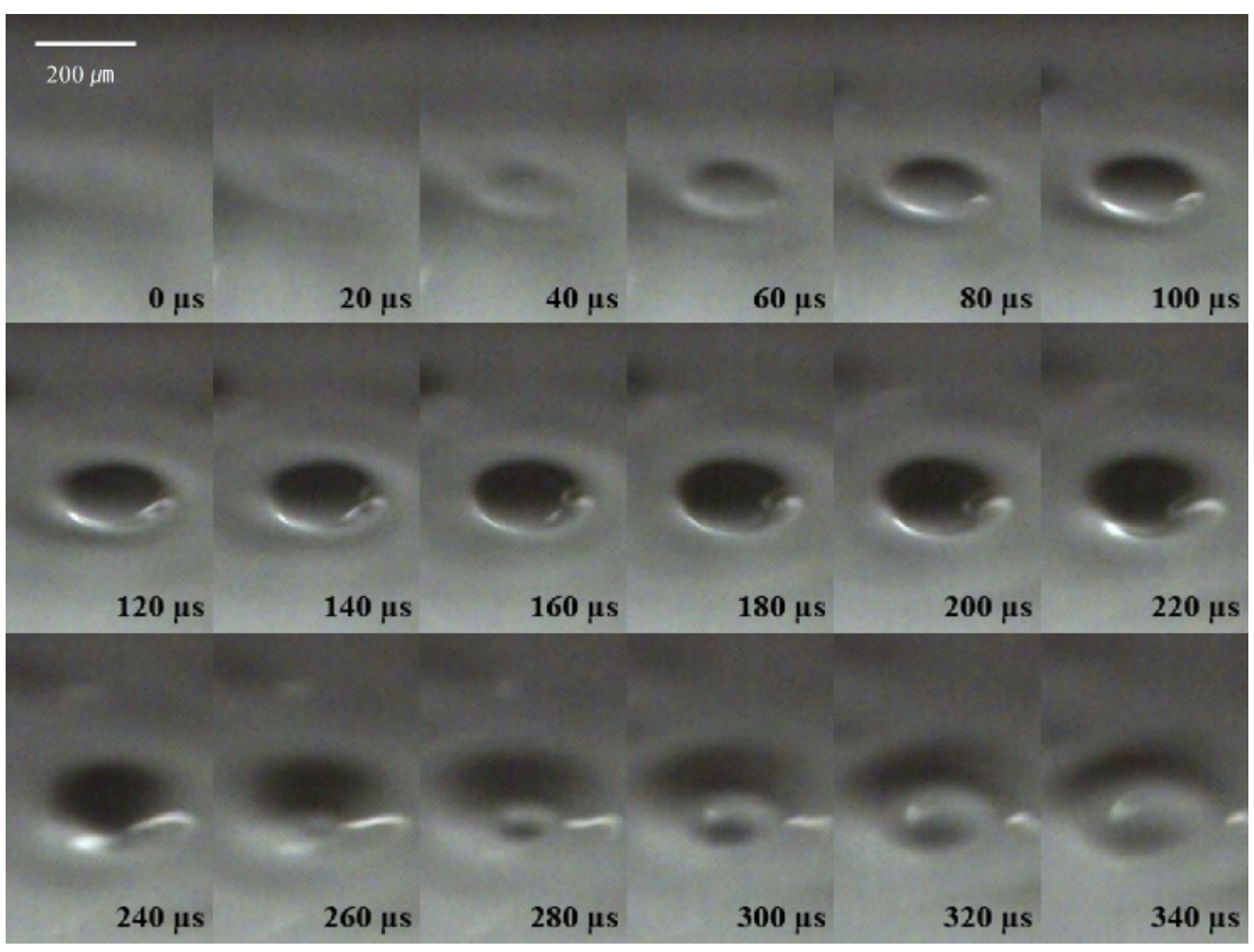

**Figure 4.** High speed imaging results for the surface deformation process. Note that concentration of NaCl solution is 20 mM and bias voltage is -540V.

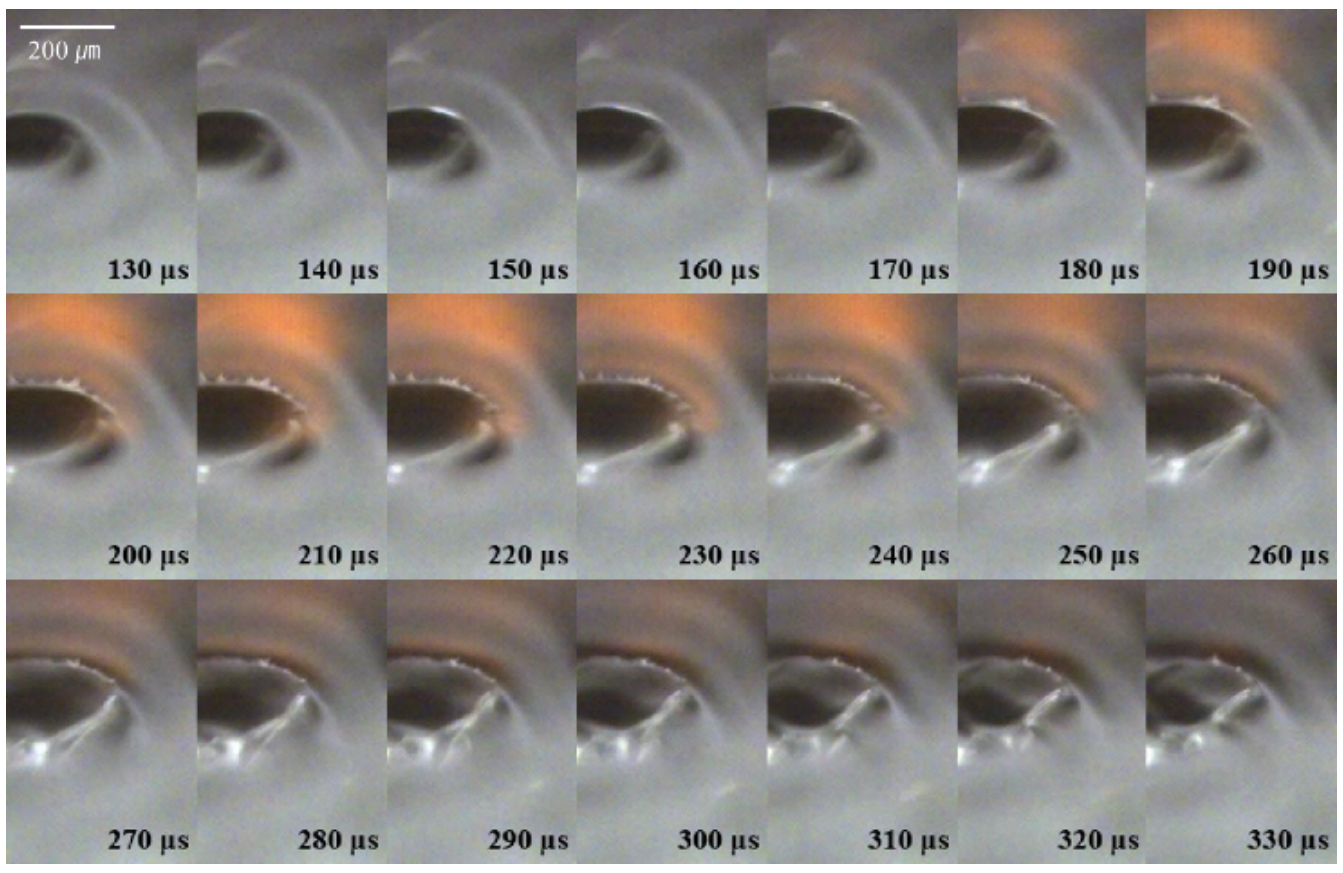

**Figure 5.** High speed imaging results for Taylor cone formation and optical Na emission. Note that concentration of NaCl solution is 100 mM and bias voltage is -540V.

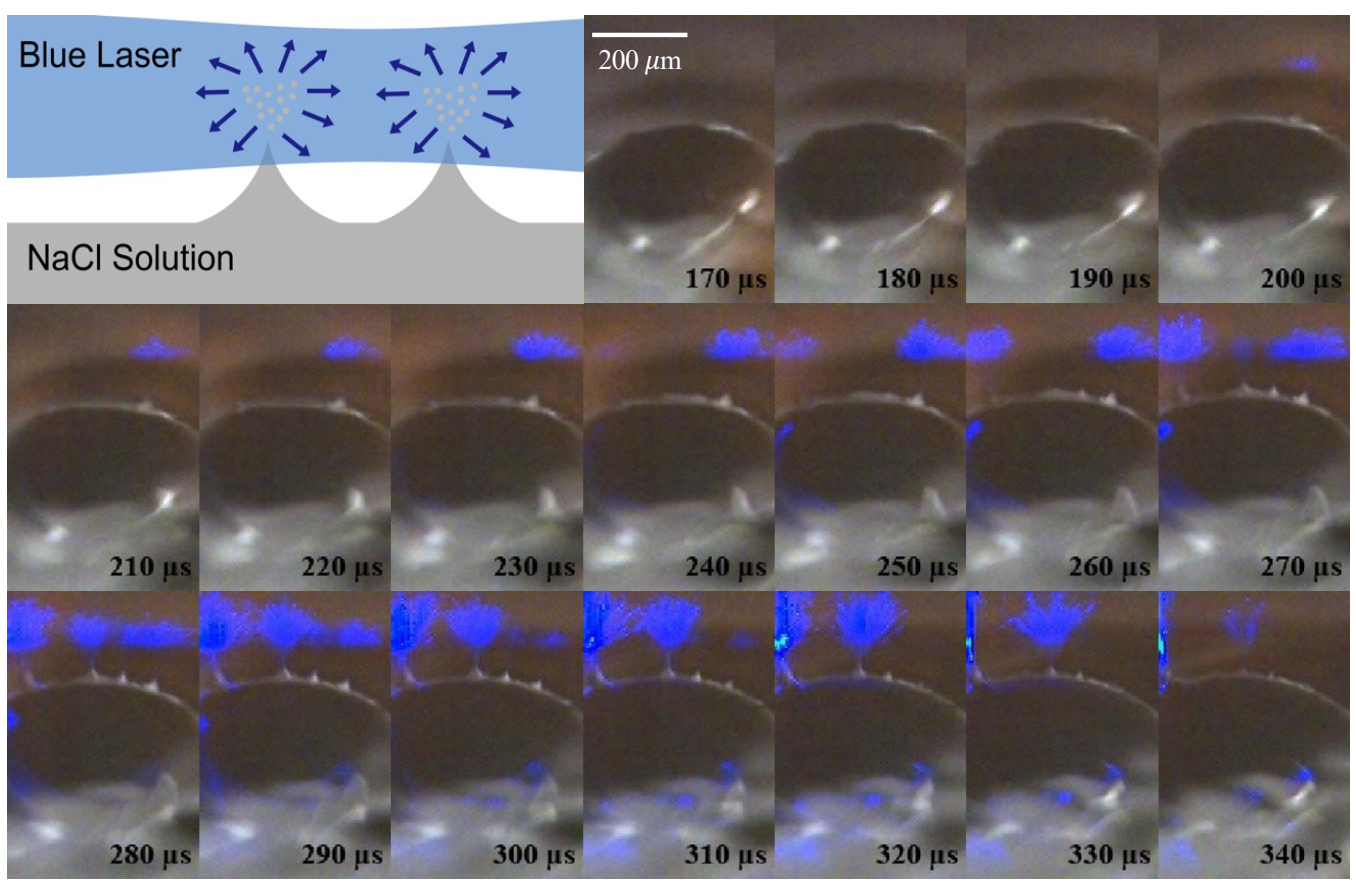

**Figure 6.** High speed imaging with Laser Mie scattered signals. The blue colored emitters corresponds to the Mie-scattered signal from liquid droplets. Note that concentration of NaCl solution is 100 mM and bias voltage is -540V.

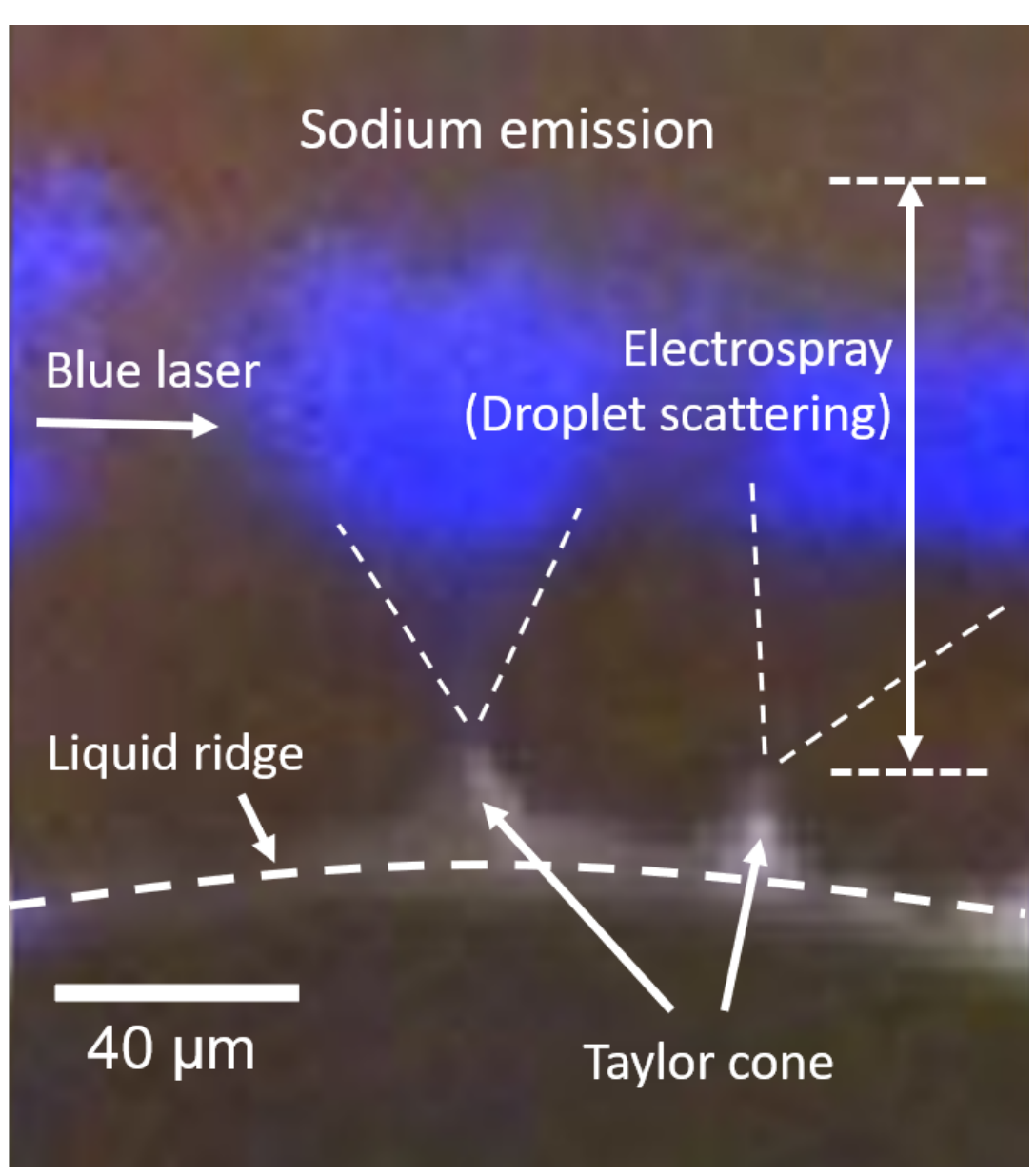

**Figure 7.** Direct visualization of Taylor cone formation and electrospray emission at the plasma–liquid interface. A Taylor cone develops from the liquid ridge beneath the localized sodium emission region. Electrosprayed droplets are visualized through scattering of a side-illuminated blue laser, revealing a spray plume extending several tens of micrometers above the liquid surface. The dashed line indicates the liquid surface profile.

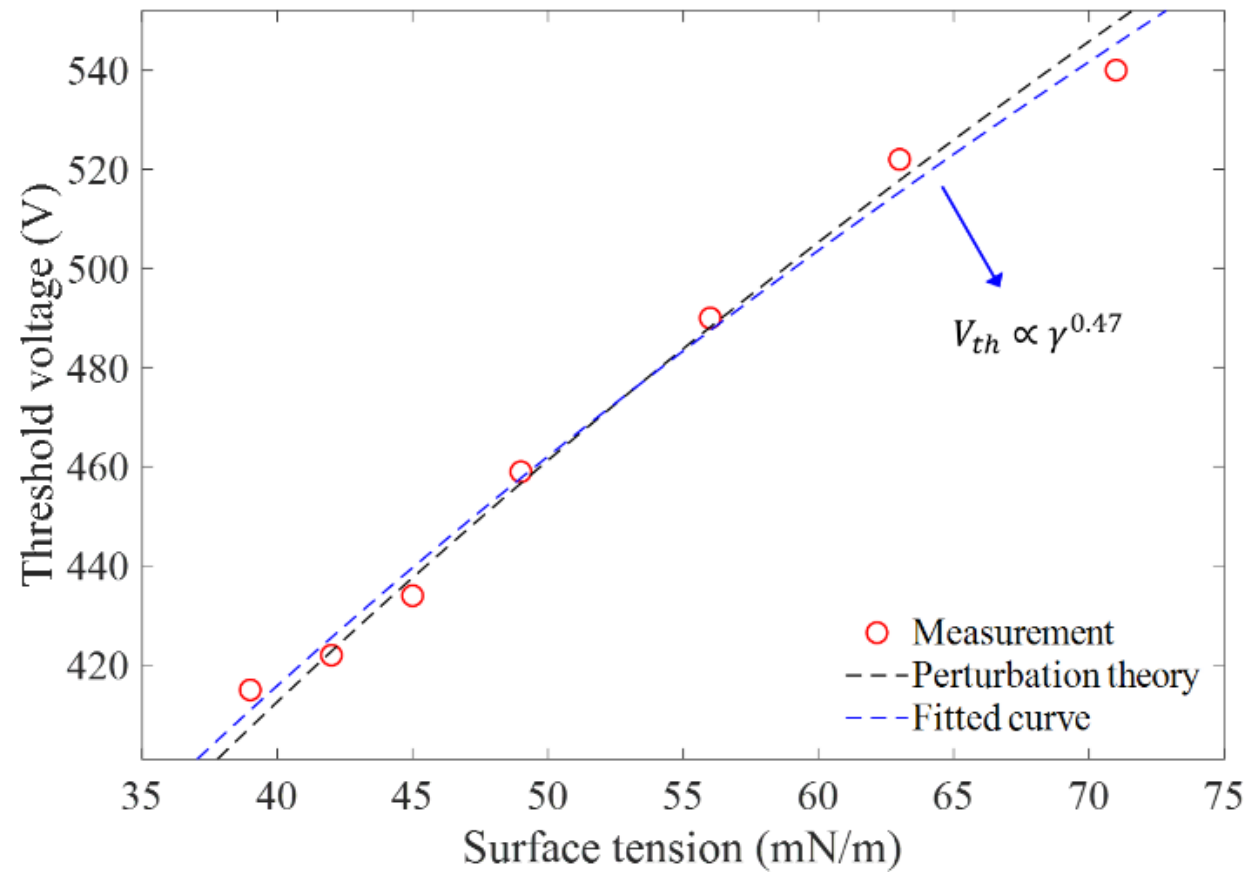

**Figure 8.** Threhold voltage requird for optical Na I emission at different surface tensions. Note that the surface tension of solutions was adjusted to be ranging from 38 to 72 mN/m by adding sodium dodecyl sulfate, while maintaining a consistent electrical conductivity (~ 200 mS/cm) [28, 29].